\documentclass[%
aps,
prl,
twocolumn,
groupedaddress,
]{revtex4-2}

\usepackage{graphicx}% Include figure files
\usepackage{dcolumn}% Align table columns on decimal point
\usepackage{bm}% bold math
\usepackage{chemformula}
\usepackage{siunitx}
\usepackage{romannum}
\usepackage{amsfonts}
\begin{document}

\preprint{APS/123-QED}

\title{\textbf{Hybrid Bound States in the Continuum beyond Diffraction Limit}
}%

\author{Ji Tong Wang}
\altaffiliation[Present address: ]{Emergent Photonics Research Centre, Department of Physics, Loughborough University, Loughborough, LE11 3TU, United Kingdom}
\email[e-mail: ]{jitong.wang@alumni.ucl.ac.uk}
\affiliation{%
  Department of Electronic and Electrical Engineering,\\
  University College London, Torrington Place, London WC1E 7JE, United Kingdom
}

\author{Nicolae C. Panoiu}
\email[e-mail: ]{n.panoiu@ucl.ac.uk}
\affiliation{%
  Department of Electronic and Electrical Engineering,\\
  University College London, Torrington Place, London WC1E 7JE, United Kingdom
}

\date{\today}% It is always \today, today,
             %  but any date may be explicitly specified

\begin{abstract}
Bound states in the continuum (BICs) have greatly impacted our ability to manipulate light-matter interaction at the nanoscale. However, in periodic structures, BICs are typically realized below the diffraction limit, thus leaving a broad spectral domains largely unexplored. Here, we introduce a new type of at-$\Gamma$ BICs of photonic crystal (PhC) slabs supporting higher diffraction orders, which we call \textit{hybrid BICs} (h-BICs), whereby symmetry protection and parameter tuning are utilized to suppress light emission in the zeroth- and higher-diffraction orders, respectively. By tuning certain structural parameters of the PhC slab, we fully characterize the dynamics of the topological structure of these h-BICs, including the generation, merging, splitting, and annihilation of circularly polarized states. We further show that the relative amount of light radiated in the first-order diffraction channels can be effectively controlled by simply breaking the $C_{4v}$ symmetry of the PhC slab. Our findings reveal a versatile approach to realize new types of BICs above the diffraction limit, and could potentially inspire new efforts towards development of novel photonic nanodevices, such as multi vortex-beam generators, frequency converters, and lasers.
\end{abstract}

%\keywords{Suggested keywords}%Use showkeys class option if keyword
                              %display desired
\maketitle

%\tableofcontents

% Introduction
\textit{Introduction}---Optical bound states in the continuum (BICs) are waves embedded in the continuous spectrum yet perfectly localized, exhibiting zero energy leakage into the far-field \cite{Hsu2016,Kang2023,Huang2023}. Their unique topological properties, manifesting as polarization vortices (\textit{V} points) in momentum space with integer topological charge \cite{Doeleman2018,Zhang2018}, together with their ability to confine light efficiently \cite{Kang2023,Wang2024}, have enabled many applications in BIC-based photonic platforms, such as sensing \cite{Tittl2018,Yesi2019,Aigner2024}, lasing \cite{Kodigala2017,Salerno2022,Contra2022,Cui2025}, chiroptical effects \cite{Gork2020,Over2021,Chen2023}, and nonlinear optics \cite{Liu2019,Koshe2019,Grom2025,Wang2025}. Generally, BICs in periodic structures can be classified into three types: symmetry-protected BICs arising from symmetry mismatch between the optical field and free-space waves \cite{Hsu2013,Koshelev2018}, Friedrich-Wintgen BICs originating from destructive interference between optical resonances \cite{Kang2021,Schi2024}, and accidental BICs formed through structural parameter tuning \cite{Hsu2013,Chen2019}. From a topological perspective \cite{Zhen2014}, controlling the polarization singularities via structural tuning provides an effective mechanism to exploit BICs. Thus, geometry tuning has been used to merge multiple BICs \cite{Jin2019,Kang2021}, to merge into a BIC a pair of circularly polarized states (\textit{C} points with half topological charge) with identical topological charge but opposite handedness \cite{Kang2025}, or to split BICs with higher-order topological charge into off-$\Gamma$ BICs \cite{Yoda2020}.

To date, nearly all reported BICs in periodic structures have been restricted to the spectral domain lying below the diffraction limit, chiefly because it has been demonstrated that at-$\Gamma$ symmetry-protected BICs of such structures cannot exist above this threshold \cite{Cerjan2021}. More specifically, BICs lying above the diffraction limit have only been achieved at off-$\Gamma$ points for very specific values of structural parameters, via merging pairs of \textit{C} points \cite{Ye2020,Hu2023}. In addition, optical excitation and light collection from such off-$\Gamma$ BICs require oblique incidence, which severely impedes their use in practical applications. Importantly, optical resonances above the diffraction limit and with high quality ($Q$)-factor are generally difficult to achieve in periodic structures \cite{Sadri2017}, which further limits the range of applications available in this spectral domain. It is apparent, therefore, that the availability of at-$\Gamma$ BICs lying beyond the diffraction limit would greatly mitigate these limitations and open up this key spectral domain to new optical devices and applications.

In this Letter, we introduce a new type of at-$\Gamma$ BICs of photonic crystal (PhC) slabs, which we call \textit{hybrid BICs} (h-BICs). These h-BICs lie beyond the diffraction limit, so that they contain zeroth- and higher-order diffraction channels. To achieve such h-BICs, the light emission in radiation channels is cancelled using two different mechanisms: the zeroth-order channel is suppressed via symmetry protection and the higher-order ones by tuning the structural parameters of the PhC. Importantly, we show that h-BICs can be achieved for a broad range of system parameters, for both TE and TM polarizations, and for PhCs with square and hexagonal lattices. Moreover, we reveal the relationship between the origin of the \textit{V} points in the higher-order diffraction channels and the topology and dynamics of \textit{C}-point pairs in the momentum space upon the variation of the geometrical parameters of the PhC. In particular, we show that the evolution of polarization singularities is governed by the conservation of the global topological charge. As a practical application of the h-BICs introduced here, we demonstrate that by breaking the $C_{4v}$ symmetry of the optical system one can readily control the relative amount of radiation emitted in the higher-order diffraction channels.

\begin{figure}[t]
  \centering
  \includegraphics[width=\columnwidth]{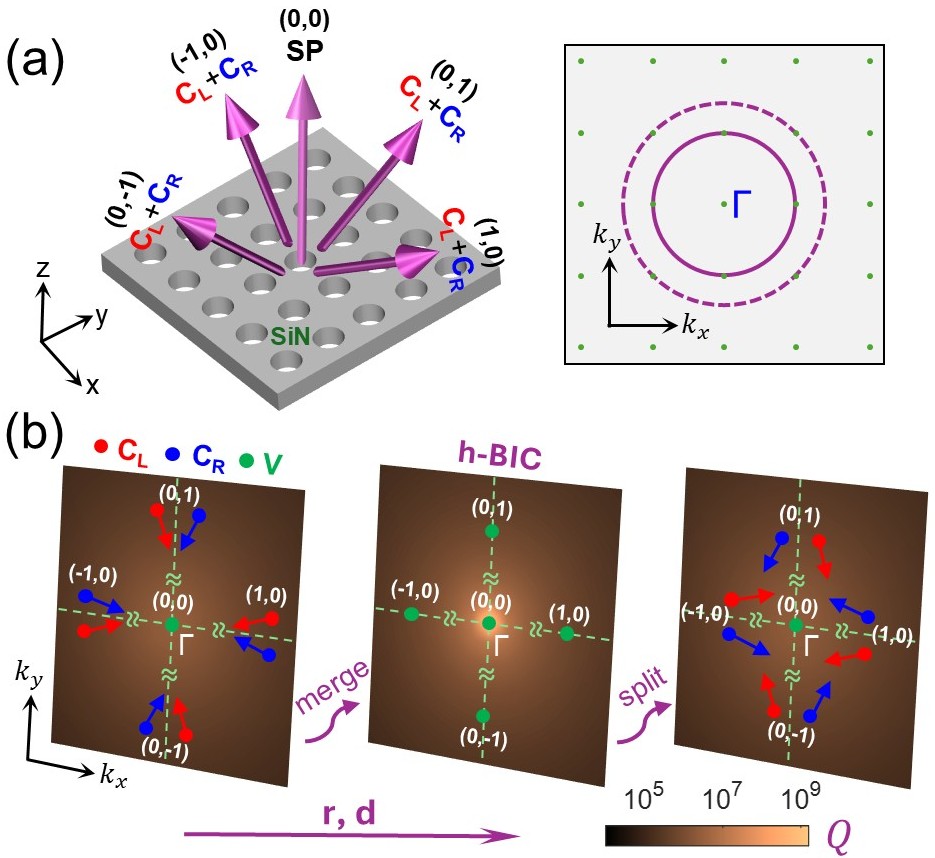}
  \caption{Concept of h-BICs. (a) Left: schematics of a PhC slab supporting zeroth- and first-diffraction orders and formation mechanism of h-BICs. $C_L$ ($C_R$) are left- (right-) handed $C$ points with the same topological charge. SP means symmetry protection. Right: reciprocal lattice space whereby the solid (dashed) circle has radius of $\omega_{I}/c$ ($\omega_{II}/c$). h-BICs can exist for $\omega>\omega_{I}$. (b) Evolution of dispersion map of $Q$ and polarization singularities upon increase of $r$ or $d$.}
  \label{fig1}
\end{figure}

\textit{Optical system configuration}---We consider a free-standing PhC slab consisting of a square array of holes, as per Fig.~\ref{fig1}(a), with out-of-plane mirror symmetry $\sigma_z$ and in-plane rotational symmetry $C_{4v}$. The PhC slab is made of SiN with $n_{SiN}\equiv n = 1.98$, has lattice constant $a=\SI{995}{\nm}$, whereas the hole radius and thickness are $r$ and $d$, respectively. At the $\Gamma$ point, the normalized frequency of the first- and second-order diffraction limits is $\tilde{\omega}_{I}=1$ and $\tilde{\omega}_{II}=\sqrt{2}$, respectively, where $\tilde{\omega}=\omega a/(2\pi c)$ is the normalized frequency, with $\omega$ and $c$ being the optical frequency and vacuum speed of light, respectively.

For at-$\Gamma$ resonances with complex frequency $\omega_c=\omega_0+i\gamma/2 $, where $\omega_0$ and $\gamma$ are the resonance frequency and leakage rate, respectively, lying between $\omega_{I}$ and $\omega_{II}$, there exist five diffraction channels: one zeroth-order channel with $\gamma_{00}$ and four first-order channels with $\gamma_{m_1m_2}$, where $(m_1,m_2)\in\mathbb{Z}^2$ and $|m_1| + |m_2| = 1$. Each pair $(m_1,m_2)$ defines an in-plane wavevector $\mathbf{k}_{\Vert,m_1m_2} = (2\pi/a)(m_1\hat{\mathbf{x}}+m_2\hat{\mathbf{y}})$ located along the high-symmetry directions $\Gamma$-$X$, cf. Fig.~\ref{fig1}(a). Due to the $C_{4v}$ symmetry of the system, at the $\Gamma$-point all four $\gamma_{m_1m_2}$ are equal. Therefore, the $Q$-factor of at-$\Gamma$ resonances is given by $Q = \omega_0/\gamma = \omega_0/(\gamma_{0}+4\gamma_{1})$, where $\gamma_{0}\equiv\gamma_{00}$ and $\gamma_{1}$ is equal to the common value of $\gamma_{m_1m_2}$.

To construct at-$\Gamma$ BICs, one must ensure that $\gamma_{0}=\gamma_{1}=0$, i.e. $Q\rightarrow\infty$. To this end, $\gamma_{0}$ can be readily cancelled by symmetry protection, since the at-$\Gamma$ resonance considered here is even under $C_{2v}^{z}$ symmetry transformation and thus is decoupled from the continuum along the surface normal direction, meaning there is a $V$ point in the (0,0) channel. By contrast, cancelling $\gamma_{1}$ requires a careful parameter tuning because symmetry protection cannot be used to suppress higher-order diffraction channels of at-$\Gamma$ resonances \cite{Cerjan2021}. Thus, as illustrated in Fig.~\ref{fig1}(b), by varying the parameters $r$ and/or $d$ one can change the locations of the $C$ points in the momentum space, enabling the simultaneous formation of a $V$ point in each of the first-order channels when pairs of $C$ points in these channels coalesce for certain parameter values, thus resulting in an h-BIC. In what follows, we discuss in-depth this tuning mechanism and the dynamics of the polarization singularities associated with the formation of h-BICs.
\begin{figure}[b!]
  \centering
  \includegraphics[width=\columnwidth]{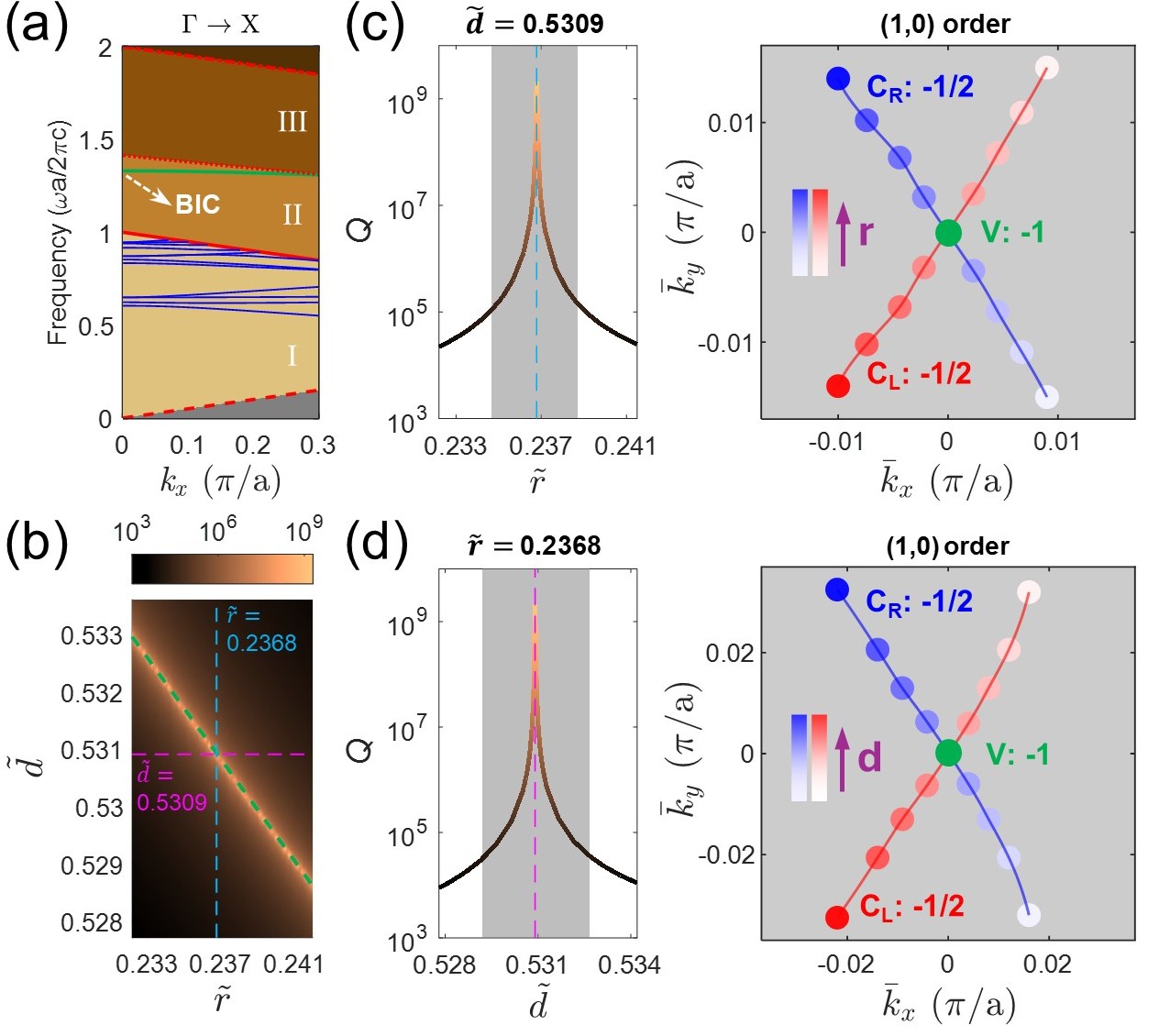}
  \caption{Characterization of h-BICs. (a) Band structure of TE-like modes. The gray region, below the light line (red dashed line), contains guided modes. Zeroth-, first-, and second-diffraction orders exist in the regions $\mathrm{I+II+III}$, $\mathrm{II+III}$, and III, respectively. Blue and green curves indicate TE-like bands and the h-BIC, respectively. (b) Dispersion map of $Q$-factor of at-$\Gamma$ h-BIC. (c) Evolution of $Q$-factor (left panel) and location of polarization singularities in the (1,0) order (right panel) vs. $r$, calculated for $\tilde{d}=0.5309$. The curves in the right panel correspond to the shaded region in the left one. (d) The same as in (c), but vs. $d$ at $\tilde{r}=0.2368$.}
  \label{fig2}
\end{figure}

\textit{Generation and physical properties of h-BICs}---We start our analysis of h-BICs by examining the band structure of the PhC slab, computed using the finite-element method \cite{COMSOL}. The mirror symmetry $\sigma_z$ allows the separation of modes into transverse electric (TE)-like and transverse magnetic (TM)-like modes. Here, we focus on the TE-like modes, the TM case being discussed in the Supplementary Material (SM) \cite{Supp}. Due to the large number of modes above $\tilde{\omega}_{I}$, we present in Fig.~\ref{fig2}(a) only the TE-like bands located in the continuum and below the line $\tilde{\omega}_{I}(\tilde{k}_x) = 1-\tilde{k}_x$, where $\tilde{k}_x=ak_x/(2\pi)$ is the normalized wavevector. It is found that the PhC slab with parameters $\tilde{r}\equiv r/a=0.2368$ and $\tilde{d}\equiv d/a=0.5309$ supports an even TE-like BIC at the $\Gamma$-point, with normalized frequency $\tilde{\omega}_{BIC}=1.331$ $(\tilde{\omega}_{II}>\tilde{\omega}_{BIC}>\tilde{\omega}_{I})$, whose dispersion curve is plotted in green in Fig.~\ref{fig2}(a). Moreover, we Fourier analyzed its optical field using the electric field distribution in an $x$-$y$ plane above the slab and determined the energy distribution among the diffraction channels (see SM \cite{Supp}). Our results reveal that at the $\Gamma$ point no energy is emitted in any of the diffraction channels, whereas finite energy leaking is observed in all five channels away from the $\Gamma$ point.

To elucidate the role of geometry tuning in the formation of h-BICs, we calculate the dependence on parameters $\tilde{r}$ and $\tilde{d}$ of the $Q$-factor, as per Fig.~\ref{fig2}(b). It can be seen that the points where $Q\rightarrow\infty$, i.e. where the BIC condition is satisfied, lie on a smooth curve in the $(\tilde{r},\tilde{d})$-space, which suggests that these h-BICs exist in a broad range of parameter space. For $\tilde{r}=0.2368$ and $\tilde{d}=0.5309$ used in Fig.~\ref{fig2}(a), the electric field profile in the middle plane of the PhC slab is even under $C_{2v}$ symmetry transformation (see SM \cite{Supp}), confirming that $\gamma_{00}=\gamma_{0}=0$. Next, we vary $\tilde{r}$ ($\tilde{d}$) at fixed $\tilde{d}=0.5309$ ($\tilde{r}=0.2368$) to further investigate the underlying physics of formation of h-BICs. As shown in Fig.~\ref{fig2}(c), $Q$ diverges at $\tilde{r}=0.2368$ (marked by a turquoise dashed line) and drops rapidly on either side of this value.

As BICs are topological $V$-points in momentum space, a key quantity characterizing these polarization singularities is the topological charge, $q$, defined as \cite{Zhen2014,Yoda2020}:
\begin{equation}
    q=\frac{1}{2\pi}\oint d \mathbf{k}_{\Vert} \cdot\nabla_{\mathbf{k}_{\Vert}}\phi(\mathbf{k}_{\Vert}),
    \label{equ1}
\end{equation}
where $\phi(\mathbf{k}_{\Vert})=\frac{1}{2} \mathrm{arg}[S_1(\mathbf{k}_{\Vert})+iS_2(\mathbf{k}_{\Vert})]$ is the polarization angle and $S_i(\mathbf{k}_{\Vert})$ are the Stokes parameters of the far-field polarization vector $\mathbf{d}(\mathbf{k}_{\Vert})$ (see SM \cite{Supp}).

Due to the $C_{4v}$ symmetry, it is enough to use just one first-order diffraction channel, say (1,0), to study the dynamics of polarization singularities upon variation of system parameters, as per Fig.~\ref{fig2}(c). Here, the shifted wavevector $(\bar{k}_x,\bar{k}_y)=(k_x-2\pi/a,k_y)$ is introduced. As the hole radius $r$ increases, a pair of $C$ points with $-1/2$ charge but opposite handedness moves towards $(\bar{k}_x,\bar{k}_y)=(0,0)$ and at $\tilde{r}=0.2368$ it merges into a $V$ point with charge $q=-1$, thus forming an at-$\Gamma$ h-BIC. When $r$ further increases, this $V$ point splits into two $C$ points. The positions of $C_L$ and $C_R$ points in the $\mathbf{k}_{\parallel}$-space define two lines that cross at $(\bar{k}_x,\bar{k}_y)=(0,0)$. Note that these two $C$ points are related by mirror symmetry $\sigma_y$, thus occurring in a pair but with opposite handedness. As shown in Fig.~\ref{fig2}(d), a similar scenario is observed if one varies $d$ at $\tilde{r}=0.2368$.
\begin{figure}[t!]
  \centering
  \includegraphics[width=\columnwidth]{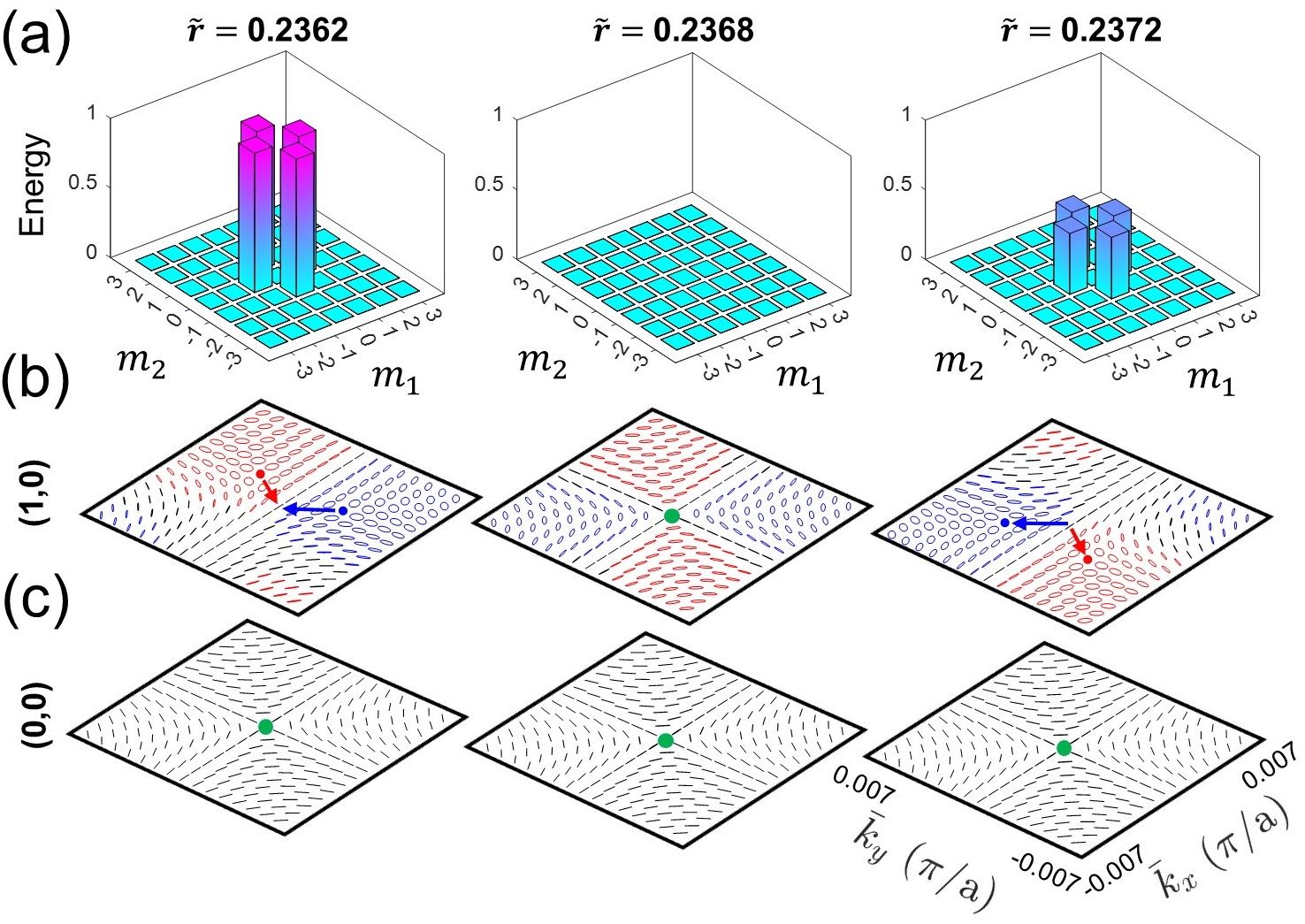}
  \caption{Fourier modal analysis and far-field polarization. (a) Energy distribution among diffraction channels, determined for different $\tilde{r}$ at fixed $\tilde{d}=0.5309$. (b), (c) Polarization maps computed for (1,0) and (0,0) diffraction channels, respectively. Arrows denote the movement direction of $C$ points as $\tilde{r}$ increases. Red (blue) indicates left (right) handedness.}
  \label{fig3}
\end{figure}

Let us now analyze the radiated energy distribution among the diffraction orders of the at-$\Gamma$ resonance, for three values of $\tilde{r}$, but for fixed $\tilde{d}$ -- see Fig.~\ref{fig3}(a). It can be observed that the energy of the (0,0) channel vanishes in all cases, whereas the four first-order channels carry an equal amount of energy that vanishes at $\tilde{r}=0.2368$. Moreover, the far-field polarization states for the $(1,0)$ and $(0,0)$ channels are illustrated in Figs.~\ref{fig3}(b) and \ref{fig3}(c), respectively (see SM \cite{Supp} for the results for all 5 channels). These plots show that as $\tilde{r}$ varies a $V$ point emerges in the $(1,0)$ channel at $\tilde{r}=0.2368$, whereas, at $\Gamma$, there is always a $V$ point with $-1$ charge in the $(0,0)$ channel due to symmetry protection. Therefore, for $\tilde{r}=0.2368$ and $\tilde{d}=0.5309$, all five channels are characterized by a topological charge of $-1$ and zero energy emission, leading to the formation of an h-BIC beyond the diffraction limit. Interestingly, the $Q$-factor decreases quadratically away from the $\Gamma$-point, $Q \propto 1/k^2_{\Vert}$, along $\Gamma\mathrm{X}$ and $\Gamma\mathrm{M}$ directions (see SM \cite{Supp}). This is so because the $Q$-factor of each channel follows $1/k^2_{\Vert}$ dependence, and thus the total $Q$-factor satisfies the same scaling law.
\begin{figure}[t!]
  \centering
  \includegraphics[width=\columnwidth]{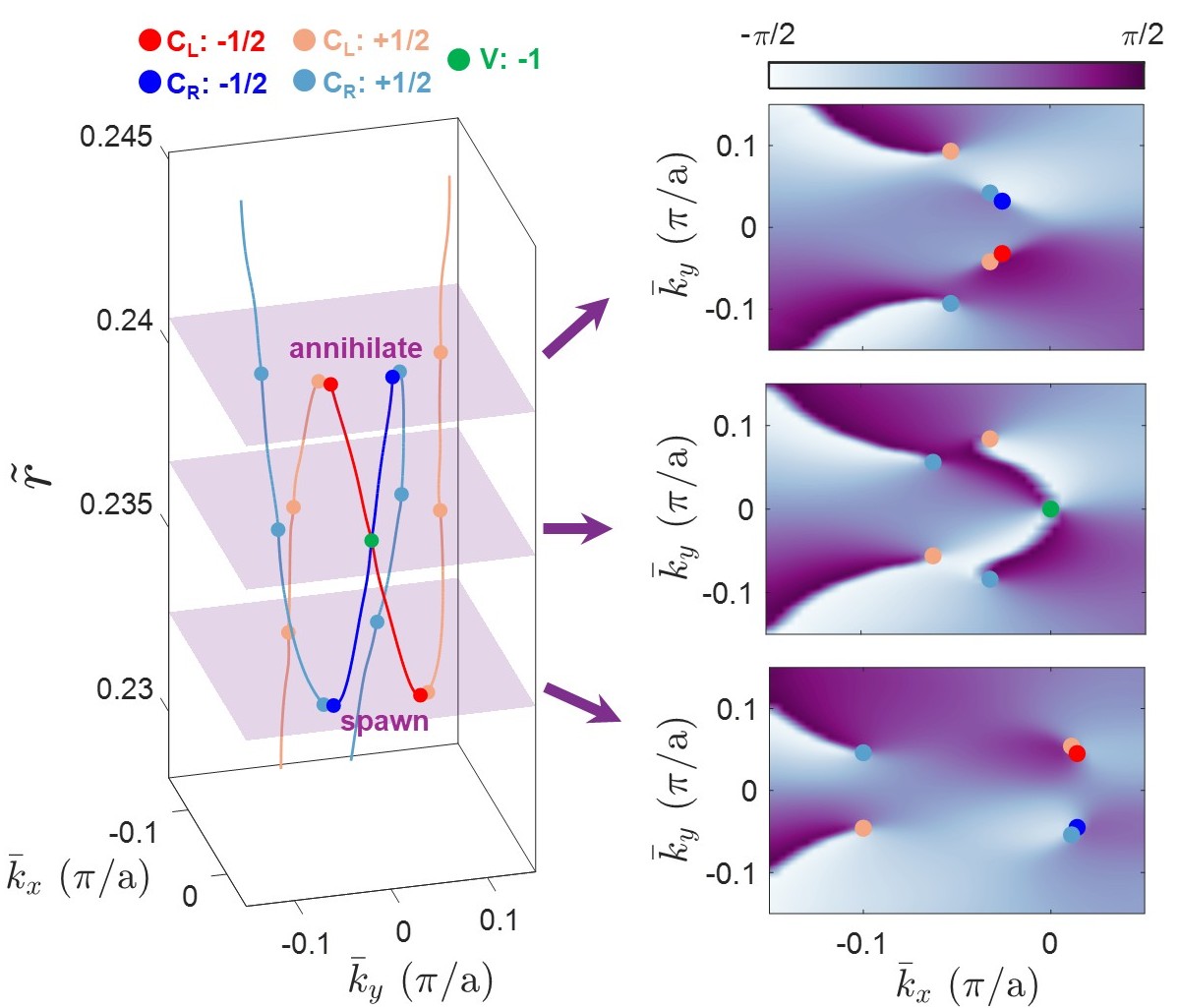}
  \caption{$\mathbf{k}_{\parallel}$-space trajectories of half-charge $C$ points for (1,0) channel, as $\tilde{r}$ is varied at $\tilde{d}=0.5309$. Red (blue) dots denote $C$ points with $-1/2$ charge and left (right) handedness. Orange (turquoise) dots denote $C$ points with $+1/2$ charge and left (right) handedness. Right panels show, from bottom to top, maps of far-field polarization angle determined for $\tilde{r}$\ =\ \numlist[list-final-separator={, and }]{0.2327;0.2368;0.2407}.}
  \label{fig4}
\end{figure}

We now seek to better understand the relation between the formation of h-BICs and the dynamics of first-order diffraction $C$ points upon variation of the system parameters. To this end, we calculated the far-field polarization states for $\tilde{r}$ ranging from \numrange{0.2281}{0.2452} at fixed $\tilde{d}=0.5309$ and observed that the system possesses several pairs of half-charge $C$ points. To characterize their dynamics, we plot in Fig.~\ref{fig4} (left panel) their trajectories in the $\mathbf{k}_{\parallel}$-space for the (1,0) radiative channel. Thus, for $\tilde{r}<0.2327$, only one pair of $C$ points with charge +1/2 and opposite handedness exist. At $\tilde{r}=0.2327$, two pairs of $C$ points spawn from states with \textit{zero} charge: one left-handed pair with opposite charge at $\bar{k}_y>0$ and one right-handed pair, again with opposite charge, at $\bar{k}_y<0$. When $\tilde{r}$ increases to \num{0.2368}, the two $C$ points with $-1/2$ charge and opposite handedness merge into a $V$ point with $-1$ charge, located at $(\bar{k}_x,\bar{k}_y)=(0,0)$, forming an h-BIC that is shown in Fig.~\ref{fig2}(c), too. Further increasing $\tilde{r}$ causes the $V$ point to split in two $C$ points with $-1/2$ charge, which subsequently annihilate the initial pair of $C$ points with $+1/2$ charge at $\tilde{r}=0.2407$. For $\tilde{r}>0.2407$, only a pair of $C$ points with $+1/2$ charge and opposite handedness exist. The corresponding maps of the polarization angle, presented in Fig.~\ref{fig4} (right panel), clearly show the change in polarization orientation around the singularities (see SM \cite{Supp} for detailed far-field polarization maps). During this evolution process, the total topological charge is conserved and equal to $q=1$.
\begin{figure}[b!]
  \centering
  \includegraphics[width=\columnwidth]{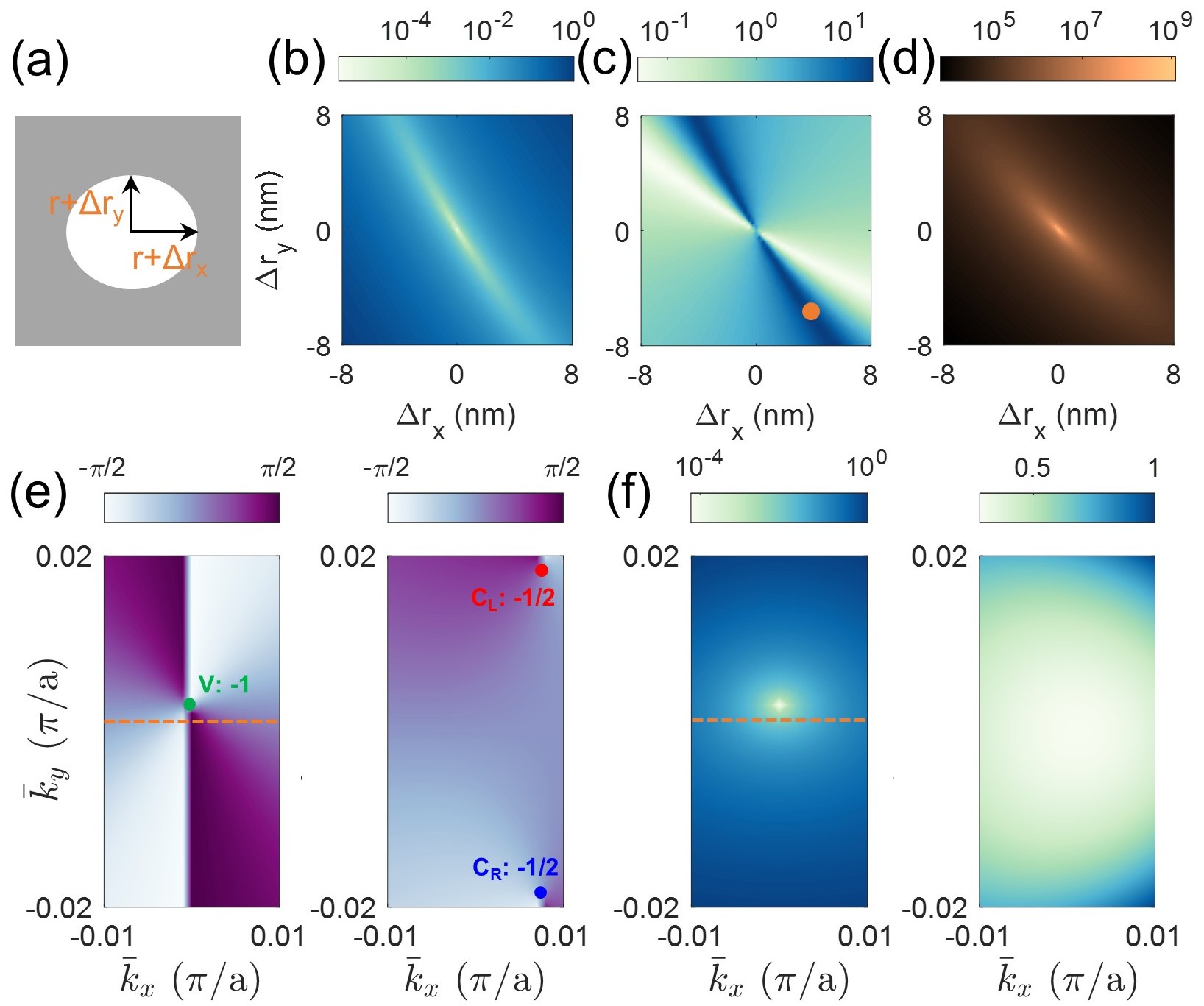}
  \caption{Engineering the radiation pattern. (a) $C_{4v}$ symmetry breaking by tuning hole radius. (b), (c), (d) Dependence of $I_{0,1}$, $I_{1,0}/I_{0,1}$, and $Q$-factor, respectively, on $\Delta r_x$ and $\Delta r_y$. The orange dot corresponds to $\Delta r_x=\SI{4}{\nm}$ and $\Delta r_y=-\SI{6}{nm}$. (e), (f) $\mathbf{k}_{\parallel}$-space maps of polarization angle and intensity $I$ for (0,1) (left) and (1,0) (right) channels, respectively. The orange lines mark $\bar{k}_y=0$.}
  \label{fig5}
\end{figure}

\textit{Applications of h-BICs}---One envisions that h-BICs containing both zeroth- and first-order diffraction channels could potentially enable a series of applications, including radiation pattern engineering, generation of multiple vortex beams, and multi-vortex beam lasers. Here, we discuss in more detail the former one and briefly suggest how the others could be implemented. Thus, to control the structure of the radiation pattern, we break the $C_{4v}$ symmetry of the PhC slab by introducing small perturbations, $\Delta r_x$ and $\Delta r_y$, of the hole radius along the $x$- and $y$-axes, respectively, as per Fig.~\ref{fig5}(a). This converts the at-$\Gamma$ h-BIC into a \textit{quasi}-h-BIC. For convenience, we now introduce a dimensionless radiation intensity defined as $I=1/Q$, which is proportional to the amount of energy emitted in a channel with $Q$-factor, $Q$. Due to the $C_{2v}$ symmetry of the PhC, at the $\Gamma$-point $I_{0,0}=0$, $I_{1,0}=I_{-1,0}$, and $I_{0,1}=I_{0,-1}$.

The distribution of $I_{0,1}$, plotted in Fig.~\ref{fig5}(b) -- see SM \cite{Supp} for the case of $I_{1,0}$, shows that it is vanishingly small at $\Delta r_x=\Delta r_y=0$, suggesting the presence of an h-BIC. When $\Delta r_x$, $\Delta r_y\neq0$, a deep trench in $I_{0,1}$ is formed and a monotonous increase of radiation away from the $\Gamma$-point is observed. Also, the distribution of the ratio $I_{1,0}/I_{0,1}$ presented in Fig.~\ref{fig5}(c) reveals a sharp contrast between its maximum and minimum values, namely 20 vs. 0.05, respectively, which proves the large tunability of the radiation pattern. In this range of $\Delta r_x$ and $\Delta r_y$, the quasi-h-BIC has $Q\gtrsim10^4$, as per Fig.~\ref{fig5}(d).

To further investigate the tunability of the radiation pattern, we select a PhC with $\Delta r_x=\SI{4}{\nm}$ and $\Delta r_y = -\SI{6}{\nm}$, corresponding to the orange dot in Fig.~\ref{fig5}(c), for which $I_{1,0}/I_{0,1}\approx20 $ (see SM \cite{Supp} for the $I_{1,0}/I_{0,1}\approx0.05$ case). As seen in Fig.~\ref{fig5}(e), the distribution of the polarization angle in the $\mathbf{k}_{\parallel}$-space indicates the presence of a singular $V$ point at $\bar{\mathbf{k}}_{\parallel,V}=(\bar{k}_x,\bar{k}_y)=(0,3)\cdot\num{e-3}\,\pi/a\neq\mathbf{0}$ in the (0,1) channel, and two singular $C$ points at $(\bar{k}_x,\bar{k}_y)=(8,\pm19)\cdot\num{e-3}\,\pi/a$ in the (1,0) channel. The corresponding radiation maps shown in Fig.~\ref{fig5}(f) confirm that $I_{0,1}=0$ at the $V$ point. Moreover, we observed that $I_{0,1}$ decays quadratically away from the $V$ point.

Our approach presented here serves as a generic method for realizing h-BICs above the diffraction limit in periodic structures. It is a versatile approach that can be used to find both TM- and TE-like h-BICs in PhC with square and hexagonal lattices (see SM \cite{Supp}), and, equally important, these ideas can be extended to h-BICs for which both first- and second-order diffraction channels can be suppressed via parameter tuning. This means that in the case of PhC with square (hexagonal) lattice they contain 9 (13) $V$ points, which can have different topological charge (see SM \cite{Supp}). These h-BICs can have immediate applications to generation of multi-vortex beams and multi-vortex beam lasers for parallel optical signal processing, as well as to advanced wavelength conversion nanodevices.

\textit{Conclusion}---In conclusion, we introduced a powerful method to construct at-$\Gamma$ hybrid BICs above the diffraction limit in PhC slabs by combining symmetry protection and parameter tuning. We reveal an intricate topology-governed evolution process involving the generation (annihilation) of a pair of $C$ points from (into) $zero$ charge states, as well as the merging (splitting) of pairwise $C$ points into (from) a $V$ point. We also demonstrate that these results are valid for PhCs with square and hexagonal lattice and for both TE and TM polarizations. Our findings significantly extend the accessible spectral range for various topological singularities, opening new avenues for high-$Q$ photonics, far-field polarization manipulation, chiroptical effects, and multiplexed diffraction control.


\begin{thebibliography}{99}

\bibitem{Hsu2016}
C. W. Hsu, B. Zhen, A. D. Stone, J. D. Joannopoulos, and M. Soljacic, Bound states in the continuum, Nat. Rev. Mater. \textbf{1}, 16048 (2016).

\bibitem{Kang2023}
M. Kang, T. Liu, C. T. Chan, and M. Xiao, Applications of bound states in the continuum in photonics, Nat. Rev. Phys. \textbf{5}, 659 (2023).

\bibitem{Huang2023}
L. Huang, L. Xu, D. A. Powell, W. J. Padilla, and A. E. Miroshnichenko, Resonant leaky modes in all-dielectric metasystems: fundamentals and applications, Phys. Rep. \textbf{1008}, 1 (2023).

\bibitem{Doeleman2018}
H. M. Doeleman, F. Monticone, W. den Hollander, A. Alu, and A. F. Koenderink, Experimental observation of a polarization vortex at an optical bound state in the continuum, Nat. Photon. \textbf{12}, 397 (2018).

\bibitem{Zhang2018}
Y. Zhang, A. Chen, W. Liu, C. W. Hsu, B. Wang, F. Guan, X. Liu, L. Shi, L. Lu, and J. Zi, Observation of polarization vortices in momentum space, Phys. Rev. Lett. \textbf{120}, 186103 (2018).

\bibitem{Wang2024}
J. Wang, P. Li, X. Zhao, Z. Qian, X. Wang, F. Wang, X. Zhou, D. Han, C. Peng, L. Shi, and J. Zi, Optical bound states in the continuum in periodic structures: mechanisms, effects, and applications, Photonics Insights \textbf{3}, R01 (2024).

\bibitem{Tittl2018}
A. Tittl, A. Leitis, M. Liu, F. Yesilkoy, D.-Y. Choi, D. N. Neshev, Y. S. Kivshar, and H. Altug, Imaging-based molecular barcoding with pixelated dielectric metasurfaces, Science \textbf{360}, 1105 (2018).

\bibitem{Yesi2019}
F. Yesilkoy, E. R. Arvelo, Y. Jahani, M. Liu, A. Tittl, V. Cevher, Y. Kivshar, and H. Altug, Ultrasensitive hyperspectral imaging and biodetection enabled by dielectric metasurfaces, Nat. Photonics \textbf{13}, 390 (2019).

\bibitem{Aigner2024}
A. Aigner, T. Weber, A. Wester, S. A. Maier, and A. Tittl, Continuous spectral and coupling-strength encoding with dual-gradient metasurfaces, Nat. Nanotechnol. \textbf{19}, 1804 (2024).

\bibitem{Kodigala2017}
A. Kodigala, T. Lepetit, Q. Gu, B. Bahari, Y. Fainman, and B. Kante, Lasing action from photonic bound states in continuum, Nature \textbf{541}, 196 (2017).

\bibitem{Salerno2022}
G. Salerno, R. Heilmann, K. Arjas, K. Aronen, J. P. Martikainen, and P. Torma, Loss-driven topological transitions in lasing, Phys. Rev. Lett. \textbf{129}, 173901 (2022).

\bibitem{Contra2022}
R. Contractor, W. Noh, W. Redjem, W. Qarony, E. Martin, S. Dhuey, A. Schwartzberg, and B. Kante, Scalable single-mode surface-emitting laser via open-Dirac singularities, Nature \textbf{608}, 692 (2022).

\bibitem{Cui2025}
J. Cui, S. Han, B. Zhu, C. Wang, Y. Chua, Q. Wang, L. Li, A. G. Davies, E. H. Linfield, and Q. J. Wang, Ultracompact multibound-state-assisted flat-band lasers, Nat. Photonics \textbf{19}, 643 (2025).

\bibitem{Gork2020}
M. V. Gorkunov, A. A. Antonov, and Y. S. Kivshar, Metasurfaces with maximum chirality empowered by bound states in the continuum, Phys. Rev. Lett. \textbf{125}, 093903 (2020).

\bibitem{Over2021}
A. Overvig, N. Yu, and A. Alu, Chiral quasi-bound states in the continuum, Phys. Rev. Lett. \textbf{126}, 073001 (2021).

\bibitem{Chen2023}
Y. Chen, H. Deng, X. Sha, W. Chen, R. Wang, Y. H. Chen, D. Wu, J. Chu, Y. S. Kivshar, S. Xiao, and C. W. Qiu, Observation of intrinsic chiral bound states in the continuum, Nature \textbf{613}, 474 (2023).

\bibitem{Koshe2019}
K. Koshelev, Y. Tang, K. Li, D.-Y. Choi, G. Li, and Y. Kivshar, Nonlinear metasurfaces governed by bound states in the continuum, ACS Photonics \textbf{6}, 1639 (2019).

\bibitem{Liu2019}
Z. Liu, Y. Xu, Y. Lin, J. Xiang, T. Feng, Q. Cao, J. Li, S. Lan, and J. Liu, High-$Q$ quasibound states in the continuum for nonlinear metasurfaces, Phys. Rev. Lett. \textbf{123}, 253901 (2019).

\bibitem{Grom2025}
D. Gromyko, J. S. Loh, J. Feng, C. W. Qiu, and L. Wu, Enabling all-to-circular polarization up-conversion by nonlinear chiral metasurfaces with rotational symmetry, Phys. Rev. Lett. \textbf{134}, 023804 (2025).

\bibitem{Wang2025}
J. T. Wang and N. C. Panoiu, Nonlinear optical metasurfaces empowered by bound-states in the continuum, Rev. Phys. \textbf{13}, 100117 (2025).

\bibitem{Hsu2013}
C. W. Hsu, B. Zhen, J. Lee, S. L. Chua, S. G. Johnson, J. D. Joannopoulos, and M. Soljacic, Observation of trapped light within the radiation continuum, Nature \textbf{499}, 188 (2013).

\bibitem{Koshelev2018}
K. Koshelev, S. Lepeshov, M. Liu, A. Bogdanov, and Y. Kivshar, Asymmetric metasurfaces with high-$Q$ resonances governed by bound states in the continuum, Phys. Rev. Lett. \textbf{121}, 193903 (2018).

\bibitem{Kang2021}
M. Kang, S. Zhang, M. Xiao, and H. Xu, Merging bound states in the continuum at off-high symmetry points, Phys. Rev. Lett. \textbf{126}, 117402 (2021).

\bibitem{Schi2024}
C. Schiattarella, S. Romano, L. Sirleto, V. Mocella, I. Rendina, V. Lanzio, F. Riminucci, A. Schwartzberg, S. Cabrini, J. Chen, L. Liang, X. Liu, and G. Zito, Directive giant upconversion by supercritical bound states in the continuum, Nature \textbf{626}, 765 (2024).

\bibitem{Chen2019}
W. Chen, Y. Chen, and W. Liu, Singularities and Poincaré indices of electromagnetic multipoles, Phys. Rev. Lett. \textbf{122}, 153907 (2019).

\bibitem{Zhen2014}
B. Zhen, C. W. Hsu, L. Lu, A. D. Stone, and M. Soljacic, Topological nature of optical bound states in the continuum, Phys. Rev. Lett. \textbf{113}, 257401 (2014).

\bibitem{Jin2019}
J. Jin, X. Yin, L. Ni, M. Soljacic, B. Zhen, and C. Peng, Topologically enabled ultrahigh-$Q$ guided resonances robust to out-of-plane scattering, Nature \textbf{574}, 501 (2019).

\bibitem{Kang2025}
M. Kang, M. Xiao, and C. T. Chan, Janus bound states in the continuum with asymmetric topological charges, Phys. Rev. Lett. \textbf{134}, 013805 (2025).

\bibitem{Yoda2020}
T. Yoda and M. Notomi, Generation and annihilation of topologically protected bound states in the continuum and circularly polarized states by symmetry breaking, Phys. Rev. Lett. \textbf{125}, 053902 (2020).

\bibitem{Cerjan2021}
A. Cerjan, C. Jorg, S. Vaidya, S. Augustine, W. A. Benalcazar, C. W. Hsu, G. von Freymann, and M. C. Rechtsman, Observation of bound states in the continuum embedded in symmetry bandgaps, Sci. Adv. \textbf{7}, eabk1117 (2021).

\bibitem{Ye2020}
W. Ye, Y. Gao, and J. Liu, Singular points of polarizations in the momentum space of photonic crystal slabs, Phys. Rev. Lett. \textbf{124}, 153904 (2020).

\bibitem{Hu2023}
P. Hu, C. Xie, Q. Song, A. Chen, H. Xiang, D. Han, and J. Zi, Bound states in the continuum based on the total internal reflection of Bloch waves, Natl. Sci. Rev. \textbf{10}, nwac043 (2023).

\bibitem{Sadri2017}
Z. F. Sadrieva, I. S. Sinev, K. L. Koshelev, A. Samusev, I. V. Iorsh, O. Takayama, R. Malureanu, A. A. Bogdanov, and A. V. Lavrinenko, Transition from optical bound states in the continuum to leaky resonances: role of substrate and roughness, ACS Photonics \textbf{4}, 723 (2017).

\bibitem{COMSOL}
COMSOL Multiphysics\textsuperscript{\textregistered}, www.comsol.com.

\bibitem{Supp}
See Supplemental Material for details of (S1) Numerical simulation methods; (S2) Bragg-diffraction orders and partition of band diagram; (S3) Definition of 2D polarization vector for topological charge calculation; (S4) $Q$-factor in a radiative channel; (S5) Fourier analysis of radiative fields; (S6) Further characterization of hybrid BICs; (S7) Symmetry properties of polarization in first-order diffraction channels; (S8) Further analysis of polarization maps; (S9) Tuning of radiation intensity in first-order diffraction channels; (S10) Additional hybrid BICs with up to first-order diffraction channels; (S11) Hybrid BICs with up to second-order diffraction channels; and (S12) Example of hybrid BICs in PhC slabs with hexagonal lattice.

\end{thebibliography}
\end{document}